# Finding citations for PubMed: A large-scale comparison between five freely available bibliographic data sources

Zhentao Liang[1,2], Jin Mao[1,2,*], Kun Lu[3], Gang Li[1,2]

**Abstract**
As an important biomedical database, PubMed provides users with free access to abstracts of its documents. However, citations between these documents need to be collected from external data sources. Although previous studies have investigated the coverage of various data sources, the quality of citations is underexplored. In response, this study compares the coverage and citation quality of five freely available data sources on 30 million PubMed documents, including OpenCitations Index of CrossRef open DOI-to-DOI citations (COCI), Dimensions, Microsoft Academic Graph (MAG), National Institutes of Health's Open Citation Collection (NIH-OCC), and Semantic Scholar Open Research Corpus (S2ORC). Three gold standards and five metrics are introduced to evaluate the correctness and completeness of citations. Our results indicate that Dimensions is the most comprehensive data source that provides references for 62.4% of PubMed documents, outperforming the official NIH-OCC dataset (56.7%). Over 90% of citation links in other data sources can also be found in Dimensions. The coverage of MAG, COCI, and S2ORC is 59.6%, 34.7%, and 23.5%, respectively. Regarding the citation quality, Dimensions and NIH-OCC achieve the best overall results. Almost all data sources have a precision higher than 90%, but their recall is much lower. All databases have better performances on recent publications than earlier ones. Meanwhile, the gaps between different data sources have diminished for the documents published in recent years. This study provides evidence for researchers to choose suitable PubMed citation sources, which is also helpful for evaluating the citation quality of free bibliographic databases.

**Introduction**

As an indispensable component of scientific publications, citation allows the author to include relevant studies in his/her own work for various motivations, including acknowledgment, criticism, persuasion, and background reading (Tahamtan and Bornmann 2018). With backward citations (also known as references), one can trace the intellectual bases of academic work (Hammarfelt 2011). On the other hand, the scientific impact and derived studies of this work can also be investigated through its forward citations (Hu et al. 2011). The importance of citation is even more prominent when it comes to the field of bibliometrics. By recording the citation relationship between publications, citation indices (e.g., Science Citation Index and Social Sciences Citation Index) have been developed and used for document retrieval and analytical purposes. Subsequently, numerous bibliometric applications have emerged, including research evaluation (Abdul-Majeed et al. 2021; Hirsch 2005), identification of research front (Small et al. 2014; Wang 2018), and topic evolution analysis (Chen et al. 2017; Han 2020). These applications serve as valuable tools for researchers and decision-makers, but they also heavily rely on citation data.

---

[1] Center for Studies of Information Resources, Wuhan University, Bayi Road #299, Wuchang District, Wuhan, 430072, Hubei, China
[2] School of Information Management, Wuhan University, Bayi Road #299, Wuchang District, Wuhan, 430072, Hubei, China
[3] School of Library and Information Studies, University of Oklahoma, Norman, OK 73019, USA
* Corresponding author: Jin Mao (danveno@163.com)

Clarivate's Web of Science (WoS) and Elsevier's Scopus are the two most comprehensive bibliographic databases that provide users with the metadata of hundreds of millions of scientific publications and citations between these documents. In fact, WoS and Scopus have been the preferred choices for many bibliometrics studies as they are recognized as authoritative and accurate data sources (Zhu and Liu 2020). However, both databases are commercial, charging a substantial cost for the access to their data. Such a paywall impedes researchers without institutional access to perform analyses on the two platforms. To alleviate this problem, the open access (OA) movement has been proposed and enacted by organizations and scientists from various disciplines. For instance, the establishment of PubMed and arXiv enables access to the abstracts and full texts of research papers free of charge. This results in accelerated scholarly communication as well as increased exposure of publications (Wang et al. 2015). It is also found that open access may promote the impact of scientific documents (Koler-Povh et al. 2014).

As one of the most comprehensive life science and biomedical databases, PubMed provides researchers with free access to the abstracts of its documents. It is an important database for literature-based discovery and other bibliometric research. During the COVID-19 pandemic, PubMed plays a critical role in grasping the latest scientific findings and supporting decision making. A number of studies have been conducted on it to help the fight against COVID-19, including drug repurposing/discovery (Mohamed et al. 2021), evolutionary path analysis (Ho and Liu 2021), and topic analysis (Zhang et al. 2021). While the PubMed documents are open to researchers, the citation relationships are not shipped with them. In other words, citation links between these documents are still "protected" behind the paywall. This poses a significant challenge to bibliometricians, who have to augment PubMed with external data sources to continue their citation-based studies (Boyack et al. 2020; Xu et al. 2020).

To rectify this, scholars urge that citation data should be recognized as a part of the commons that are freely and legally available for sharing (Shotton 2013). New academic products that involve citation data have also been developed and open to the public, including Dimensions, Microsoft Academic Graph (MAG), Semantic Scholar, and PubMed Central (PMC). National Institutes of Health (NIH) has also launched its official project to construct an open citation collection for PubMed documents (Ian Hutchins et al. 2019). With such a wide range of free bibliographic data sources, a critical research problem has emerged: **Which is the most suitable free data source for providing PubMed with citation relationships?**

Related studies have been conducted to analyze the differences between academic databases, primarily focusing on the coverage of documents (Harzing 2016), journals (Singh et al. 2021), scholars (Harzing and Alakangas 2017), and institutions (Hug and Brändle 2017). However, besides the coverage, it is also essential to evaluate the correctness and other aspects of the quality of the databases. Missing and false citations may cause problems in bibliometric practices, especially for the studies that are sensitive to citation relationships, e.g., main path analysis and link prediction. In addition, most previous studies compared databases based on relatively small or biased samples. They did not take PubMed as a research object, either.

To fill these gaps, we study the quality of citation relationships provided by five freely available data sources based on the 2020 version of PubMed Baseline[4], which contains over 30 million publications but most of the citations are not given. The free bibliographic data sources are compared pairwise and against gold standards (PubMed, Scopus, and WoS) in terms of various evaluation metrics. Different from previous studies, we avoid selection bias by adopting a complete set of documents and retrieving their backward citations from each database. In addition to coverage, we define other metrics to investigate the data quality, including accuracy, precision, and recall. Through diachronic analyses, our research also shows how citation quality changes over time. Our study is beneficial to researchers to select the proper free citation sources, especially those without institutional access. Moreover, this study is primarily based on open datasets, which increases the reproducibility and validity of our results. The extracted PMID-to-PMID citations in this study are available on Zenodo[5].

This paper is organized as follows. In Related work, we briefly review related work on academic databases comparison. The data collection, processing, and evaluation methods are described in the Methodology section, following which, we present the results. In Discussion, we elaborate the implications, limitations, and future work of this study. Finally, we conclude our research in the Conclusions.

**Related work**

To investigate the differences between academic databases, scholars first performed bibliometric analyses based on small samples of documents. Harzing (2016) compared the publication and citation coverage of Microsoft Academic Graph, Google Scholar, WoS, and Scopus. The comparison was based on her 124 publications and forward citations, showing that Google Scholar and Microsoft Academic Graph had a larger coverage than WoS and Scopus. Later, Harzing and Alakangas (2017) expanded the sample to the publications authored by 145 scholars from the same university. The result also confirmed that the two databases not only covered more publications but also had higher citation counts of these documents. In a similar vein, research of this type starts with a controlled set of documents, which usually consists of article lists maintained by academics or institutions, and iteratively searches for these documents as well as their citations in candidate databases (Bar-Ilan 2010; de Winter et al. 2014; Harzing 2019; Hug et al. 2017; Hug and Brändle 2017). Although these studies have provided insight into the coverage of different data sources, their analyses were based on small and different sets of documents. Therefore, results may vary when using other samples, undermining the generality and comparability of the research.

Recently, more studies have focused on larger and more comprehensive sets of articles. Thelwall (2017) sampled 172,752 articles from 29 large journals for analysis. As he paid special attention to the disciplinary differences of citation counts, the selected journals covered 26 Scopus broad fields. Martín-Martín et al. (2018) used 2,515 highly-cited articles displayed in Google Scholar's Classic Papers as seed documents. A total of 2,448,055 forward citations of these documents were extracted from three databases, covering almost all subject areas. In addition to disciplinary differences in citation coverage, they also revealed characteristics of unique citations in each database. According to their results, Google Scholar had a significant advantage in indexing non-journal and non-English citations. In the follow-up research, they included three more data sources and performed a pairwise comparison (Martín-Martín et al. 2020). Google Scholar still ranked first in terms of the coverage of citations, followed by Microsoft Academic Graph. While a larger and systematic sample increases the validity of these studies, three limitations remain as follows: (1) Samples used in previous research were biased toward highly-cited, classical articles and those published in large journals. (2) Most studies focused only on the coverage of databases. Given that citations provided by the databases may contain errors (Van Eck and Waltman 2017), additional quality measurements are needed when comparing different data sources. (3) Previous studies mainly collected forward citations of the seed documents to construct their experimental dataset. This is an effective approach to enlarge the sample, but the dynamic nature of forward citations also makes it difficult to investigate the correctness and completeness of data.

---

[4] ftp://ftp.ncbi.nlm.nih.gov/pubmed/baseline
[5] https://doi.org/10.5281/zenodo.5184461

Some researchers have realized these limitations and attempted to perform the comparison more systematically. In particular, the complete lists of documents covered by the databases have been used to eliminate the selection bias (Mongeon and Paul-Hus 2016; Visser et al. 2021). Special attention has also been paid to the correctness of citation relationships (Haunschild et al. 2018; Van Eck and Waltman 2017). The most relevant study of our research is Visser et al.'s work (Visser et al. 2021) on the large-scale comparison between five databases. They obtained the metadata of all documents from each dataset and analyzed the coverage in terms of various breakdown criteria, including publication year, document type, and language. In addition, they studied the overlap of citations between Scopus and other data sources, from which the reason for incomplete/incorrect citations was also investigated manually, such as citations towards unpublished documents and the "secondary version" of documents (preprints or proceedings). However, the main focus was still on the publication coverage of databases, and limited attention was paid to citations and corresponding evaluation metrics.

In this study, we investigate the quality of PubMed-to-PubMed citations provided by five free bibliographic data sources based on the complete set of PubMed documents. We define various quality measurements, including coverage, accuracy, precision, as well as recall, and compare data sources against three gold standards. Temporal changes of these quality metrics are also investigated. Our research provides practical references for selecting the suitable free database to augment the PubMed dataset. In addition, this study is helpful to the scientists who are interested in performing bibliometric analyses based on open/freely available data sources, by providing the knowledge of how accurate the citation links are and how the quality changes through time.

**Methodology**

Fig. 1 shows the overall framework of our research, which consists of four stages. The first step is to collect the PubMed Baseline and extract the identifier and metadata of each document. This results in a baseline dataset, whose citations are retrieved from external databases later. Next, PubMed-to-PubMed citations are collected from five freely available data sources and matched with the documents in our baseline. In the third step, a subset of documents is sampled from the baseline dataset. Their backward citations are retrieved from three authoritative databases, composing the gold standards used to evaluate the data sources. Finally, we analyze the overlap and correlation between these databases based on the complete PubMed dataset. Evaluation metrics are also developed to measure the data quality against gold standards. In addition, we perform diachronic analyses to study the temporal changes in the citation quality. It should be noticed that we intend to investigate the differences in citation quality for documents published in different years, rather than the longitudinal differences for versions of data sources released at different times. Details of these stages are described in the following subsections.

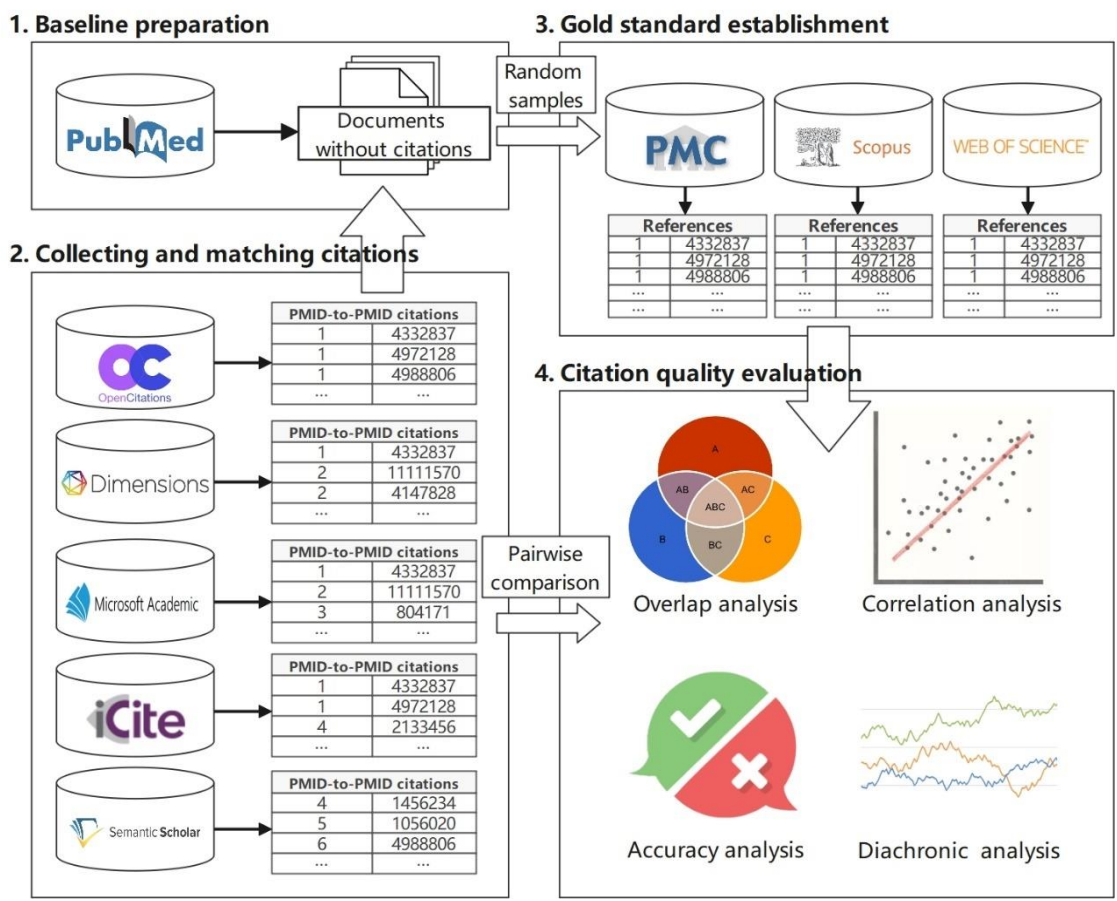

**Fig. 1**. Research framework of citation sources comparison.

**Baseline preparation**

As we focus on the citation links provided by different data sources, a set of documents without citation relationships should be defined beforehand. PubMed Baseline is an annual snapshot of PubMed that offers a complete list and metadata of documents, while the citation relationships of most of the documents are absent. We downloaded the PubMed 2020 Baseline dataset in October 2020, which contains 30 million PubMed documents published by 2019. The dataset was delivered in XML format and we extracted the metadata of all documents through Python scripts. The digital object identifier (DOI) and PubMed ID (PMID) of each document were parsed to match with other data sources. The major function of the PubMed Baseline is to provide a common platform to compare different citation sources.

**Collecting and matching citations**

In the second step, we collected the backward citations (i.e., the references) of documents in the baseline from five free bibliographic databases: OpenCitations Index of CrossRef open DOI-to-DOI citations (COCI), Dimensions, Microsoft Academic Graph (MAG), National Institutes of Health's Open Citation Collection (NIH-OCC), and Semantic Scholar's Open Research Corpus (S2ORC).

We collected references instead of forward citations because the reference list of an article is fixed after publishment, which makes it easier to examine the correctness and completeness. In addition, by collecting the references of these documents, their forward citations can also be found within the dataset. The collecting and matching procedures of each data source are described below.

*COCI.* COCI contains all the citations that are specified by the open references to DOI-identified works present in Crossref (Heibi et al. 2019). We downloaded the September 2020 Dump of COCI from OpenCitations' official website. It contained 733 million DOI-to-DOI citation links between 59 million publications (OpenCitations, 2020). By matching the DOIs with those in our baseline, we obtained 258 million referencing relationships of 11 million PubMed documents. During the peer-review process of this paper, COCI has integrated references from Elsevier in an important release (OpenCitations, 2021). We measured its performance separately (denoted as *COCI.Updated*) to compare it with the previous release.

*Dimensions.* Dimensions is a database with over 105 million publications and offers no-cost access to researchers (Herzog et al. 2020). Through the applications programming interface (API), we retrieved the references of all PubMed documents by their DOIs or PMIDs in November 2020. The DOIs and PMIDs of the references were provided if available. By matching the DOIs and PMIDs with those in our baseline, we obtained 509 million referencing relationships of 19 million PubMed documents.

*MAG.* As one of the largest academic databases, MAG contains over 240 million publications of various types (Sinha et al. 2015). We acquired the October 2020 version of MAG and extracted DOI-to-DOI citation links from it. Although PMIDs were also provided as the second-class attribute of MAG articles, we found a substantial number of articles were assigned with wrong PMIDs. For papers without DOIs, we matched them to the PubMed database jointly by the last name of the first author, International Standard Serial Number (ISSN) of the venue, publication year, volume number, and the begin page number. A MAG document was matched to a PubMed document only if they had the same values of all five fields. As a result, we obtained 480 million referencing relationships of 18 million PubMed documents.

*NIH-OCC.* To solve the problem that a large number of citations are not open, NIH started to build its own open citation collection in 2019 (Ian Hutchins et al. 2019). NIH-OCC provides native PMID-to-PMID citations between PubMed articles. We downloaded the September 2020 version of NIH-OCC (iCite et al. 2020). By matching the PMIDs with those in our baseline, we obtained 468 million referencing relationships of 17 million PubMed documents.

*S2ORC.* S2ORC is a general-purpose corpus freely available for researchers. It contains over 136 million publications and 467 million citation relationships between them (Lo et al. 2020). PMIDs and DOIs are provided if available. We downloaded the July 2020 version of S2ORC and matched the PMIDs and DOIs with those in our baseline. Finally, we obtained 183 million referencing relationships of 7 million PubMed documents from S2ORC.

**Gold standard establishment**

To evaluate the quality of citation provided by the five data sources, the actual reference lists of PubMed documents should be obtained as gold standards. We established three gold standards in this study. The first gold standard consists of citation relationships included in the PubMed Baseline, which are provided by the publishers or have full-text articles archived in PMC. It contains 174 million references of 5.5 million PubMed publications (~19%). As we focused on finding citations within the PubMed Baseline, references without PMIDs were dropped. We treated citation relationships of the first gold standard as the most authentic since they were shipped with the PubMed Baseline. On the other hand, we also expected that all data sources would achieve better results on this gold standard as the data may be directly integrated into these data sources.

In addition to the first gold standard, we also collected citation relationships from Scopus and WoS as our second and third gold standards. We chose Scopus and WoS since they are the most established bibliographic databases, which allow us to compare freely available data sources against the commercial veterans. Although the problems of missing and incorrect citations have also been found in both Scopus and WoS, it is still beneficial to investigate to what extent free data sources resemble them. We expected that the performance of free data sources would be lower on the second and third gold standards. However, the performances on the second and third gold standards are more helpful to judge the overall citation quality of free data sources as they represent a more general population.

Because we do not have access to the full Scopus and WoS databases, a representative set of documents was sampled from the PubMed Baseline and their references were collected from the two databases. This sample consists of 50,000 documents randomly sampled over the last 50 years (1970-2019), from which 1,000 documents were selected each year. The publication types were restricted to journal article and review to ensure the existence of references. Publications whose references are available in the PubMed Baseline are not selected. To collect data from Scopus, we queried the PMID and DOI of each document in our sample and exported the metadata of their references in CSV format. The Scopus API was not used because we found that the metadata of many references was missing, while they were properly supplied through the export function on the Scopus website. A similar procedure was applied to the WoS Core Collection. WoS allows exporting all metadata of the documents as well as their references in batches of up to 500.

After collecting data from Scopus and WoS, the references need to be matched with the documents in the PubMed Baseline. A reference that meets any of the following criteria were considered a match to the PubMed publication.
(1) Sharing the same PMID or DOI with a document in the PubMed Baseline.
(2) Sharing the same venue, volume, publication year, and start page with a document in the PubMed Baseline.
(3) Sharing the same title and publication year with only one document in the PubMed Baseline.

Finally, we collected 42,803 (85.6%) documents from Scopus and 977,813 references were matched in the PubMed Baseline. The figures for WoS were 37,897 (75.8%) documents and 851,521 PubMed references. Documents that were not found in the databases or had no PubMed references were not included in the two gold standards.

**Citation quality evaluation**

As reported by previous research, the problems of missing and incorrect citations exist in various bibliographical databases (Van Eck and Waltman 2017; Visser et al. 2021). For instance, in the NIH-OCC dataset, we found document PMID-23487520 published in 2013 was incorrectly reported as the reference of document PMID-15224180 published in 2004. Similar errors were also found in the reference lists of the documents PMID-8702918, PMID-12917354, and PMID-15818467. Therefore, it is necessary to understand the quality of data sources before using them in bibliometric research. While most of the comparative studies focused only on the coverage, this study also pays attention to other aspects of the citation quality. We evaluate the correctness and completeness of citation data provided by different data sources with the metrics below.

*Coverage.* It measures the proportion of PubMed documents whose references can be found in the free data source, regardless of the correctness and completeness of references. The coverage does not count the situation that the metadata of a document can be found but not its references. The coverage of a data source is calculated as follows:

$$Coverage_{DB} = \frac{count(doc_{DB} \cap doc_{PubMed})}{count(doc_{PubMed})} \tag{1}$$

*Precision.* It measures the extent to which the references provided by the free data source are correct, compared with the gold standard. We first calculated the precision of the references on every single document:

$$Precision_{doc} = \frac{TP}{TP + FP} \tag{2}$$

where *TP* denotes the references of a document provided by the data source, which are also reported as true references in the gold standard (i.e., true positive). *FP* denotes the references provided by the data source but does not exist in the gold standard (i.e., false positive). The precision of the data source is calculated as the average of document-level reference precision:

$$Precision_{DB} = \frac{1}{n} \sum_{doc=1}^{n} Precision_{doc} \tag{3}$$

It should be noted that only documents covered by the specific data source were included in the calculation process of precision. Therefore, *n* equals to the number of covered documents.

*Recall.* It measures the extent to which the references provided by the free data source are complete, compared with the gold standard. The recall of the references was first calculated for every document:

$$Recall_{doc} = \frac{TP}{TP + FN} \tag{4}$$

where *TP* has the same meaning as formula (2). *FN* denotes the references that are reported as true references of a document by the gold standard, but the relationships are not found in the free data source (i.e., false negatives). The recall of the data source is calculated as the average of document-level reference recall:

$$Recall_{DB} = \frac{1}{n} \sum_{doc=1}^{n} Recall_{doc} \tag{5}$$

Similarly, only documents covered by the specific data source were included in the calculation process of recall.

***F1-score.*** It is a measure of data quality combining both precision and recall. We adopted the macro version of F1-score of a data source in this study, which is the average document-level F1-score:

$$F1_{doc} = 2 \times \frac{Precision_{doc} \times Recall_{doc}}{Precision_{doc} + Recall_{doc}} \quad (6)$$

$$F1_{DB} = \frac{1}{n} \sum_{doc=1}^{n} F1_{doc} \quad (7)$$

***Accuracy.*** It measures the proportion of PubMed documents whose references are correctly provided by the free data source, without the problems of incorrect and missing references (formula 8). Only the documents covered by the specific data source were included in the calculation process of precision.

$$Accuracy_{DB} = \frac{count(doc_{correct})}{count(doc_{DB} \cap doc_{PubMed})} \quad (8)$$

where $doc_{correct}$ denotes documents whose reference lists are correctly and completely provided by the data source. In other words, the F1-scores of these documents equal to 1.

## Results

### Descriptive statistics

We first present the descriptive statistics of the coverage of five data sources. As Table 1 shows, Dimensions has the highest coverage that 62.4% of documents in the PubMed 2020 Baseline can find at least one reference in this data source. NIH-OCC has a lower coverage but the most references per document, while MAG has a similar coverage but fewer references on average. On the other hand, the coverage of COCI and S2ORC are substantially lower, so as the average references. The integration of Elsevier's references significantly increased the coverage of COCI.Updated by 50% with the average number of references almost unchanged. By combining the data from all five databases, 66.9% of PubMed documents can find their references and the average number of references per document is 27.3. It is worth noting that the coverage presented in Table 1 is relatively conservative since not all types of PubMed documents should have references. There are 28,374,243 articles and reviews in the PubMed 2020 Baseline. By taking this number as the denominator, we obtained the adjusted coverage for each data source (column 4, Table 1).

**Table 1** The coverage of five bibliographic data sources on the PubMed Baseline.

| Database | Publications | Coverage | Adjusted coverage | Referencing relationships* | Avg. references |
|---|---|---|---|---|---|
| COCI | 10,563,315 | 34.7% | 37.2% | 257,638,206 | 24.4 |
| COCI.Updated | 15,349,346 | 50.5% | 54.1% | 374,332,023 | 24.4 |
| Dimensions | **18,995,173** | **62.4%** | **66.9%** | **508,742,050** | 26.8 |
| MAG | 18,136,034 | 59.6% | 63.9% | 479,947,694 | 26.5 |
| NIH-OCC | 17,248,884 | 56.7% | 60.8% | 467,922,262 | **27.1** |
| S2ORC | 7,145,784 | 23.5% | 25.2% | 182,653,647 | 25.6 |
| Combined result** | 20,339,956 | 66.9% | 71.7% | 555,460,052 | 27.3 |

\* A distinct referencing relationship is identified between one citing publication and one cited publication.
\*\* Duplicate citing publications and referencing relationships are merged.

In terms of the change of coverage over time, Dimensions remains the most comprehensive data source from 1900 to 2019, followed by NIH-OCC (Fig. 2). The coverage of MAG is higher than NIH-OCC for documents published in 1997-2011 and 2018-2019, while it ranks third in other time periods. COCI and S2ORC have the lowest coverage during the whole period. The updated version of COCI has a larger coverage than the September 2020 release since 1945. It even surpasses MAG after 2018. In addition, an abnormal decrease is observed in the coverage of NIH-OCC and S2ORC in 2017. As shown in the inset plot, an increasing proportion of PubMed documents can find references in free data sources. The sudden drop in 1945 is caused by the sharp increase of indexed PubMed documents.

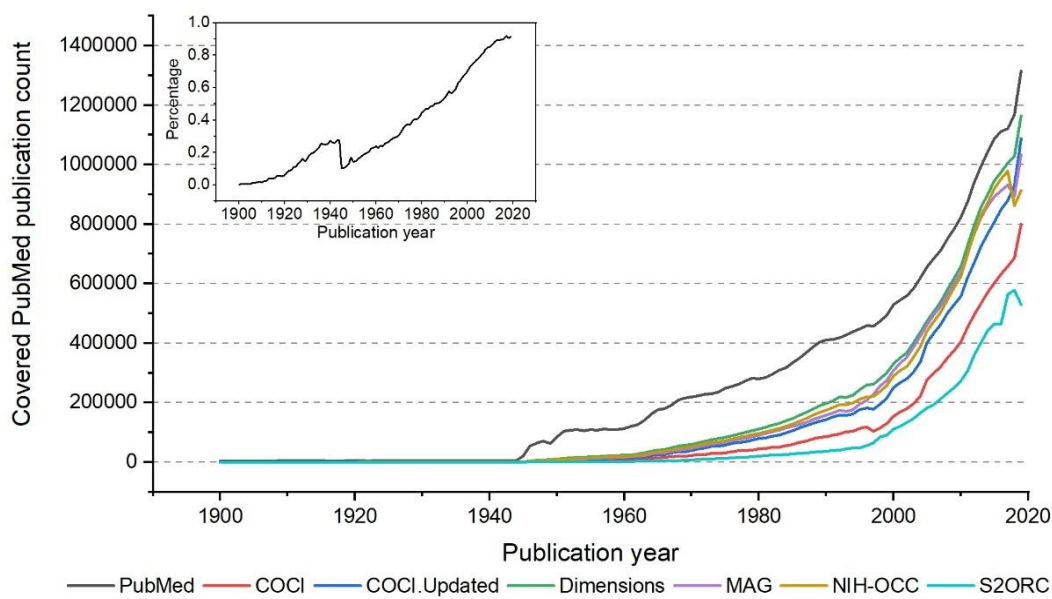

**Fig. 2**. The temporal change of coverage of each data source. The black line in the major graph represents the annual distribution of PubMed documents. The inset plot shows the percentage of PubMed documents whose references can be found in any one of the five data sources. PubMed documents published before 1900 are not included as their references cannot be found in all data sources.

**Overlap analysis**

Moreover, we investigated the document-level and referencing-relationship-level overlap among the five databases (Fig. 3). We used UpSet plots (Lex et al. 2014) to visualize the results, in which the lower area shows the exclusive combinations of data sources with their sizes in percentage above. It is shown that 22.6% of the documents and 19.2% of the referencing relationships are shared by all data sources. MAG possesses the largest number of exclusive documents (2.9%), followed by Dimensions (2.7%), whereas most of the documents in COCI can also be found in other data sources. Surprisingly, while S2ORC ranks last in terms of document coverage, it has the third-largest number of exclusive documents (1.2%). By contrast, though NIH-OCC has the second-largest document coverage, it has relatively fewer exclusive documents (< 1.0%). Regarding referencing relationships, the number of exclusive relationships is the largest in MAG (4.2%), followed by Dimensions (2.1%), S2ORC (1.5%), NIH-OCC (0.5%), and COCI (~0.0%). Documents that can be found in all databases except S2ORC constitute the largest proportion in Fig 3.a (27.0%). Similar characteristic exists in the overlap of referencing relationships (Fig 3.b).

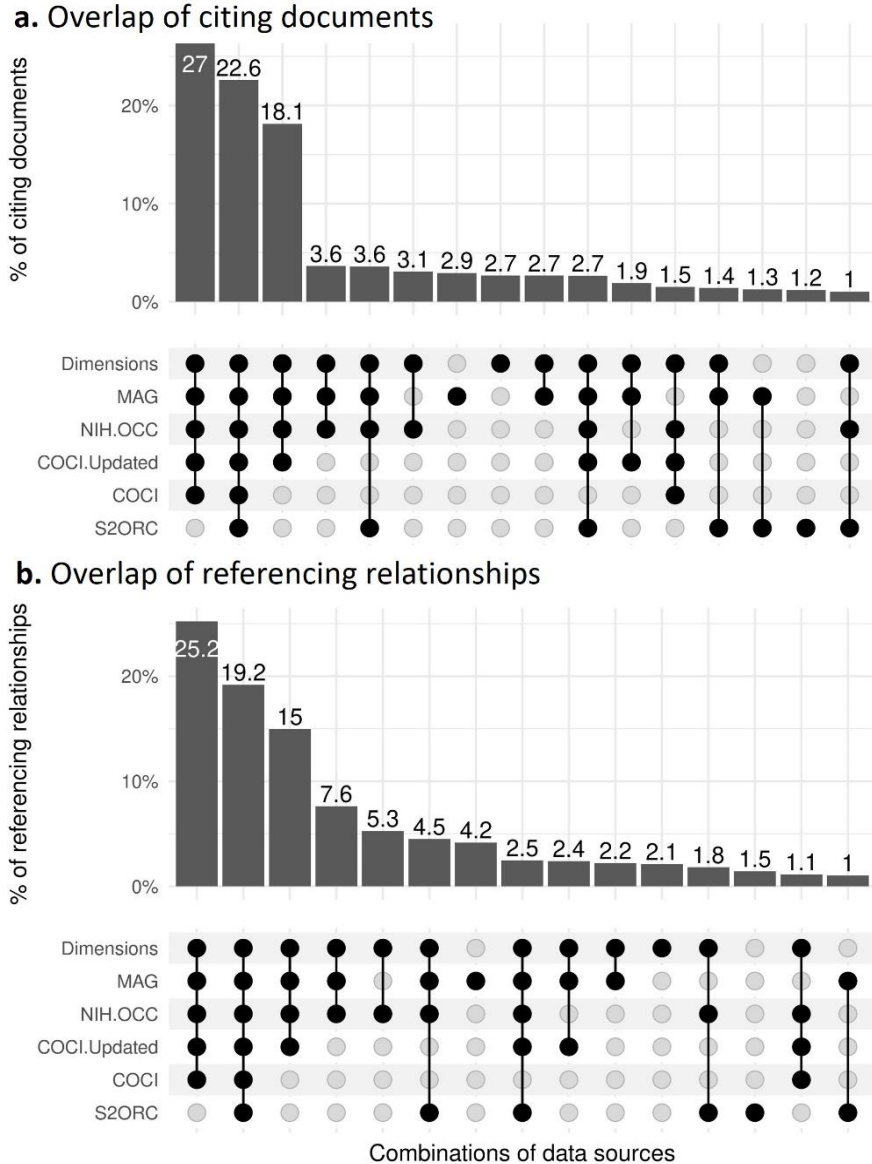

**Fig. 3**. Overlap of documents (a) and referencing relationships (b) among the five data sources. Combinations with < 1% of elements are omitted.

To understand the overlapping relationship more in-depth, we compared the documents and referencing relationship of the five databases in a pairwise way (Table 2 and 3). Again, Dimensions is the most comprehensive data source that over 90% of documents from other free bibliographic databases can be found in it. It contains almost all (99.49%) PubMed documents and referencing relationships in COCI. As an official project supported by the National Institutes of Health Office of Portfolio Analysis (OPA), NIH-OCC contains over 84% of documents and referencing relationships provided by other free data sources. Dimensions is not an NIH-supported project, but it still outperforms NIH-OCC on the coverage of PubMed. This suggests that Dimensions is a competitive free data source to use in bibliometric research. Similarly, MAG also has a larger coverage of PubMed documents (89.16%) and referencing relationships (86.41%) than NIH-OCC. Nonetheless, all data sources except NIH-OCC have significantly lower coverage of referencing relationships, compared with the coverage of citing documents. This may suggest possible missing citations problems and we further analyze them in the following sections.

**Table 2** Pairwise document overlap of five data sources.

|  | COCI | COCI.Updated | Dimensions | MAG | NIH-OCC | S2ORC |
|---|---|---|---|---|---|---|
| Overall | 51.93% | 75.46% | **93.39%** | 89.16% | 84.80% | 35.13% |
| COCI |  | **100.00%** | 99.49% | 96.34% | 99.46% | 44.44% |
| COCI.Updated | 68.82% |  | 99.37% | 96.95% | 96.35% | 34.52% |
| Dimensions | 55.33% | 80.30% |  | 90.00% | 89.78% | 34.64% |
| MAG | 56.11% | 82.05% | **94.27%** |  | 87.82% | 36.02% |
| NIH-OCC | 60.91% | 85.74% | **98.87%** | 92.34% |  | 36.04% |
| S2ORC | 65.70% | 74.16% | **92.09%** | 91.42% | 87.00% |  |

**Table 3** Pairwise referencing relationships overlap of five data sources.

|  | COCI | COCI.Updated | Dimensions | MAG | NIH-OCC | S2ORC |
|---|---|---|---|---|---|---|
| Overall | 46.38% | 67.39% | **91.59%** | 86.41% | 84.24% | 32.88% |
| COCI |  | **100.00%** | 99.49% | 96.69% | 99.30% | 42.32% |
| COCI.Upadted | 68.83% |  | **99.08%** | 96.94% | 94.82% | 33.37% |
| Dimensions | 50.38% | 72.91% |  | 87.48% | 90.53% | 32.81% |
| MAG | 51.91% | 75.61% | **92.73%** |  | 86.45% | 33.34% |
| NIH-OCC | 54.68% | 75.85% | **98.43%** | 88.67% |  | 34.09% |
| S2ORC | 59.69% | 68.38% | **91.37%** | 87.60% | 87.33% |  |

**Correlation analysis**

We also investigated how similar were citation counts of the cited documents shared by all data sources. The citation counts were obtained based on the referencing relationships within the same database. As the citation counts of documents are not normally distributed (Thelwall 2016), the Spearman correlation was calculated for each pair of data sources (Table 4). The highest correlation coefficient (0.99) is observed between Dimensions and MAG, as well as between Dimensions and NIH-OCC. With the lowest coverage, S2ORC presents a relatively low but similar correlation (0.84) with other databases. All correlation coefficients are above 0.8, suggesting the citation ranks of publications have high consistency across data sources.

**Table 4** Spearman correlation between the five data sources.

|  | COCI | COCI.Updated | Dimensions | MAG | NIH-OCC | S2ORC |
|---|---|---|---|---|---|---|
| COCI | 1.00 |  |  |  |  |  |
| COCI.Updated | 0.96 | 1.00 |  |  |  |  |
| Dimensions | 0.93 | 0.97 | 1.00 |  |  |  |
| MAG | 0.92 | 0.97 | 0.99 | 1.00 |  |  |
| NIH-OCC | 0.94 | 0.97 | 0.99 | 0.98 | 1.00 |  |
| S2ORC | 0.84 | 0.82 | 0.84 | 0.84 | 0.84 | 1.00 |

Note: 13,390,210 cited documents shared by all data sources were included in the calculation. All correlations are significant at the 0.01 level (2-tailed).

**Quality of the citations**

In addition to coverage, overlap, and correlation of citation counts, more evidence is needed to evaluate the citations provided by the free data sources. One of the most important questions is how accurate and complete these citation relationships are. The quality of the citations from free data sources is defined as the extent to which they resemble citations in the gold standards. To this end, we established three gold standards and five measurements for evaluation purposes (see the Methodology section). References provided by free data sources were compared against the true references in the gold standards. Table 5 presents descriptive statistics of the three gold standards. The PubMed Baseline gold standard is the largest and has the greatest number of average references per document. The Scopus and WoS gold standards are smaller and have a similar number of average references.

Table 5  Descriptive statistics of the gold standards.

| Gold standard | Publications | Referencing relationships* | Avg. references |
|---|---|---|---|
| PubMed Baseline | 5,512,064 | 175,958,177 | 31.9 |
| Scopus | 42,803 | 977,813 | 22.8 |
| Web of Science | 37,897 | 851,521 | 22.5 |

* A distinct referencing relationship is identified by one citing publication and one cited publication.

### Evaluation on the PubMed Baseline gold standard

Table 6 presents the evaluation results of five bibliographic data sources on the first gold standard, which consists of referencing relationships extracted from the PubMed 2020 Baseline. Since the National Library of Medicine (NLM) is one of the contributors of NIH-OCC, it is not surprising that NIH-OCC covers all documents in this gold standard. In addition, NIH-OCC achieves the best results in terms of all evaluation metrics. As a database not specifically for PubMed documents, Dimensions achieves extraordinary results that are only slightly behind NIH-OCC. Both databases provide correct and complete references for over 80% of documents in the first gold standard. One possible reason for the close performances of Dimensions and NIH-OCC is that they share similar upstream data providers (e.g., Crossref, NLM, and some partner publishers). However, they also have different internal data processing pipelines that may introduce disparity in their final datasets.

On the other hand, while MAG performs relatively well in terms of coverage (87.36%), it suffers from the problem of low accuracy (27.73%). We found that 91% and 39% of the error cases are due to incomplete (recall < 1.0) and incorrect (precision < 1.0) references, while 31% have both problems. Similar patterns were observed in other data sources, indicating that recall is a more serious problem than precision for these databases. By incorporating high-quality references from Elsevier, the updated version of COCI significantly increases its coverage with even higher precision and recall. S2ORC is inferior to other data sources in terms of all evaluation metrics.

Table 6  Performance of the five data sources on the PubMed Baseline gold standard | N=5,512,064.

|  | COCI | COCI.Updated | Dimensions | MAG | NIH-OCC | S2ORC |
|---|---|---|---|---|---|---|
| Coverage | 65.57% | 72.29% | 99.98% | 87.36% | **100.00%** | 63.03% |
| Precision | 99.87% | 99.87% | 99.60% | 97.87% | **99.90%** | 97.66% |
| Recall | 84.69% | 85.18% | 98.80% | 90.80% | **98.99%** | 79.00% |
| F1-score | 90.59% | 90.95% | 99.07% | 93.37% | **99.34%** | 86.27% |
| Accuracy | 15.67% | 15.60% | 81.55% | 27.73% | **89.08%** | 5.86% |

Note: Only documents covered by the specific data source were included in the calculation of precision, recall, and F1-score.

We further analyzed the annual changes in the five metrics. Fig. 4 shows that NIH-OCC has a coverage of 100% throughout the whole period, whereas Dimensions also has extremely similar coverage. There is an upward trend in the coverage of MAG, COCI, and S2ORC between 1970 and 2019. In particular, the coverage of MAG and COCI rise dramatically for documents published after 1995 and 2005, respectively. Also, the difference between the updated and previous version of COCI is most prominent after 2005.

Fig. 5 presents the other four metrics of each data source. It is evident that the performance of each database increases on recent publications. Among them, NIH-OCC and Dimensions make the greatest progress in terms of accuracy (Fig. 5a), while the precision of MAG and S2ORC increases most significantly (Fig. 5c). The performances of two versions of COCI on these metrics are close. In addition, NIH-OCC and Dimensions have a similar and extraordinary performance during the whole period, compared with other data sources. The gaps between different data sources diminish for documents published in recent years.

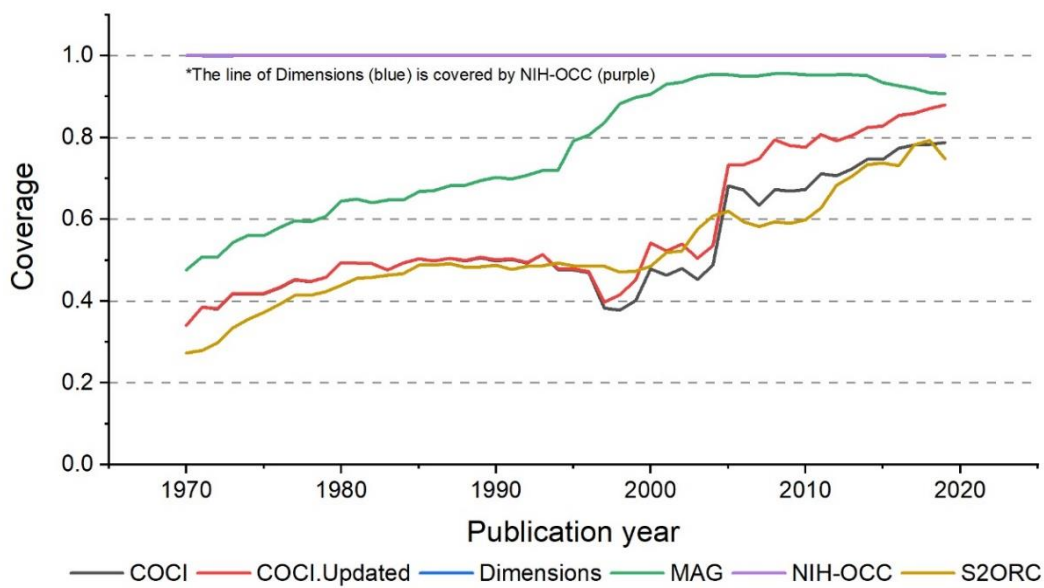

**Fig. 4**. Temporal changes in the coverage of different data sources on the PubMed Baseline gold standard.

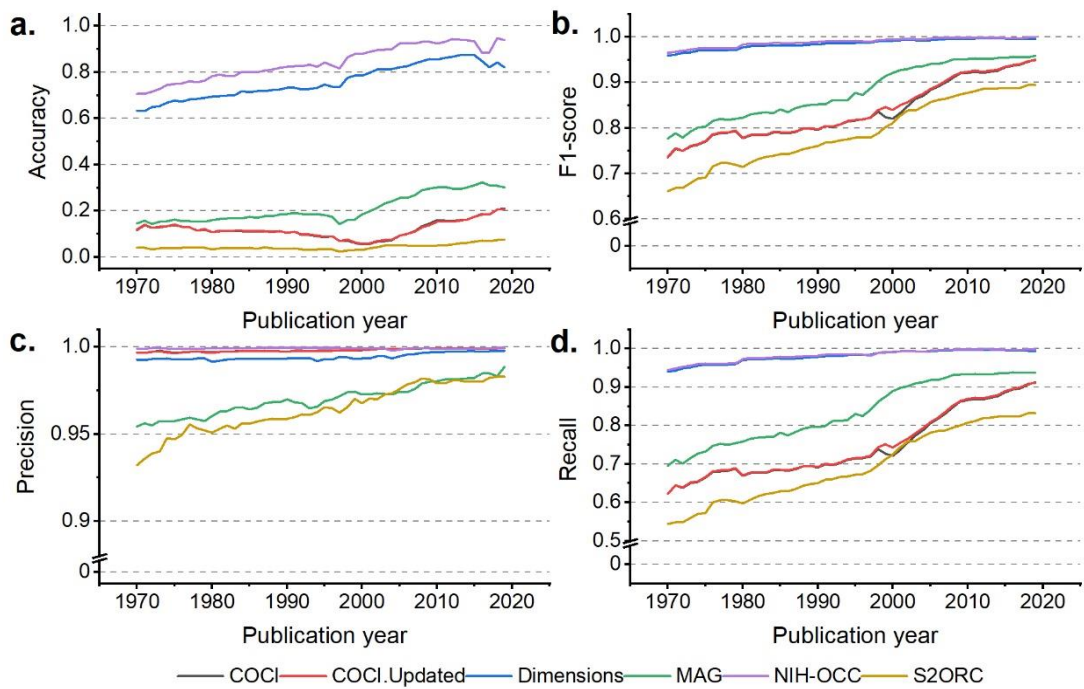

**Fig. 5**. Temporal changes in the accuracy, precision, recall, and F1-score of different data sources on the PubMed Baseline gold standard.

### Evaluation on the Scopus gold standard

Regarding the Scopus gold standard, all data sources experience a drop in performance, compared with those on the first gold standard (Table 7). It is reasonable because the references of documents in this gold standard cannot be exported directly from PubMed. By representing a more general population of documents, the Scopus gold standard can yield more informative results than the first gold standard. Intriguingly, Dimensions outperforms NIH-OCC in terms of coverage (90.53% vs. 84.10%), recall (93.85% vs. 93.44%), and accuracy (52.49% vs. 44.70%). The precision (97.70%) and F1-score (94.78%) of Dimensions are only slightly behind those of NIH-OCC (97.73% and 94.92%). This confirms that Dimensions is a competitive data source for providing citations of PubMed documents. The updated version of COCI achieves the highest precision performance. In addition, it is obvious that all data sources have a precision of over 95% but a much lower recall. MAG has the second largest coverage but lower accuracy due to serious incomplete reference problem.

**Table 7**   Performance of the five data sources on the Scopus gold standard | N=42,803.

|  | COCI | COCI.Updated | Dimensions | MAG | NIH-OCC | S2ORC |
|---|---|---|---|---|---|---|
| Coverage | 50.59% | 80.47% | **90.53%** | 86.87% | 84.10% | 20.30% |
| Precision | 97.72% | **98.05%** | 97.70% | 96.00% | 97.73% | 95.99% |
| Recall | 80.11% | 78.86% | **93.85%** | 86.40% | 93.44% | 72.56% |
| F1-score | 86.36% | 85.88% | 94.78% | 89.82% | **94.92%** | 80.85% |
| Accuracy | 22.38% | 18.87% | **52.49%** | 23.18% | 44.70% | 6.41% |

Note: Only documents covered by the specific data source were included in the calculation of precision, recall, F1-score.

Fig. 6 and 7 demonstrate the temporal performance changes of the free bibliographic databases on the Scopus gold standard. Dimensions has the largest coverage during the whole period, followed by MAG and NIH-OCC. The coverage of COCI and S2ORC is much lower. It is worth noting that the updated version of COCI has a much larger coverage than the previous version throughout this period, but their gap diminishes significantly after 2018. In general, all data sources cover more PubMed documents over time, while COCI experienced a sudden rise in 2018. In addition, Dimensions always provides the most accurate references on the Scopus gold standard, followed by NIH-OCC (Fig. 7a). It is interesting that COCI provided more accurate references for documents published between 1985 and 1996, then its accuracy dropped to 10% for those in 1997 and gradually increased afterward. The precision, recall, and F1-score of all data sources increase on more recent publications. Dimensions and NIH-OCC achieve similar recall rate and F1-score, so do MAG and COCI (Fig. 7b and 7d).

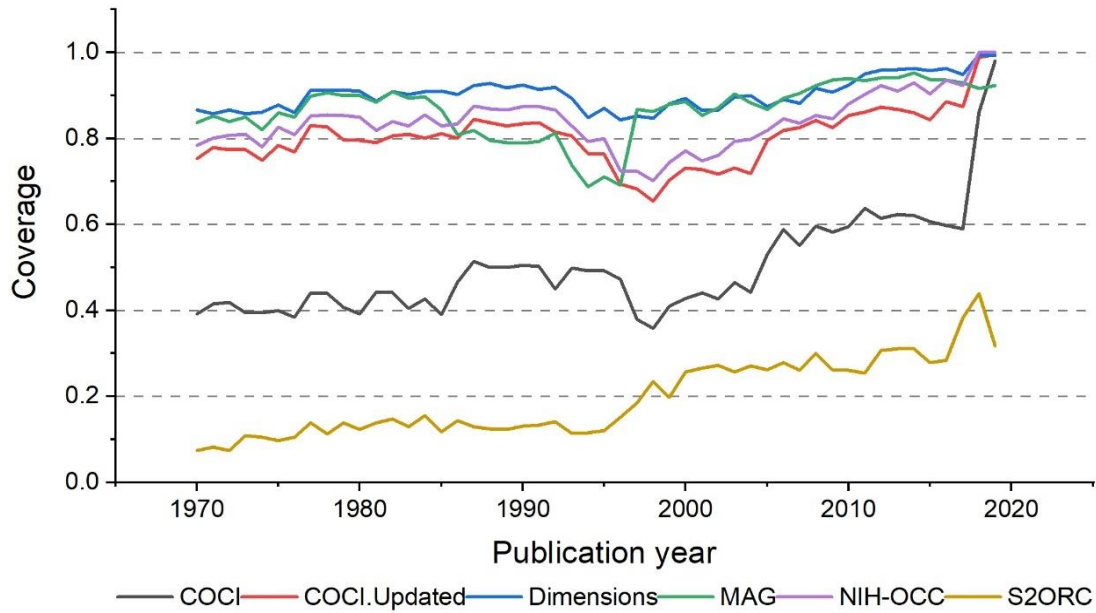

**Fig. 6**. Temporal changes in the coverage of different data sources on the Scopus gold standard.

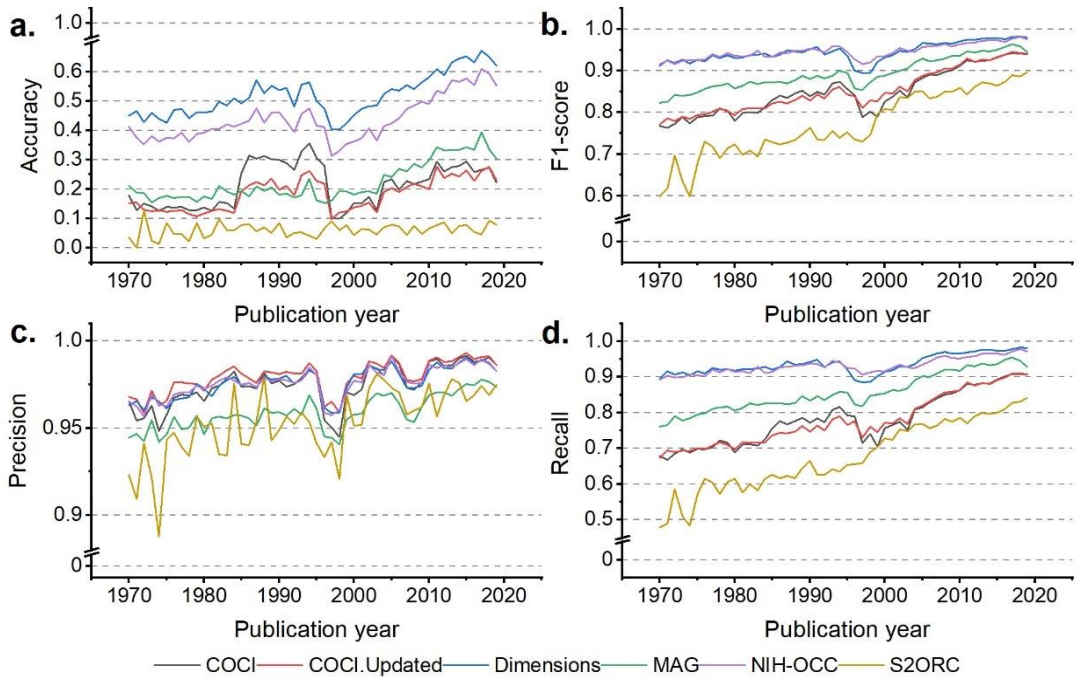

**Fig. 7**. Temporal changes in the accuracy, precision, recall, and F1-score of different data sources on the Scopus gold standard.

## Evaluation on the Web of Science gold standard

As a complement to the Scopus gold standard, we collected the references of the same samples from WoS to construct the third gold standard. Precision and accuracy of all data sources decrease significantly on the WoS gold standard, while the coverage, recall, and F1-score are less affected (Table 8). Nonetheless, the relative performance of data sources remains stable. Dimensions has the largest coverage (87.14%), followed by MAG (84.11%), NIH-OCC (79.17%), COCI.Updated (75.72%), COCI (48.76%), and S2ORC (21.24%). NIH-OCC has the highest recall (95.06%) and F1-score (92.19%), and its accuracy (26.39%) is similar to that of Dimensions (26.94%).

COCI.Updated provides the most precise (93.44%) references on this gold standard but its coverage is smaller. S2ORC is still inferior to other data sources in terms of all evaluation metrics.

**Table 8** Performance of the five data sources on the WoS gold standard | N=37,897.

|  | COCI | COCI.Updated | Dimensions | MAG | NIH-OCC | S2ORC |
| --- | --- | --- | --- | --- | --- | --- |
| Coverage | 48.76% | 75.72% | **87.14%** | 84.11% | 79.17% | 21.24% |
| Precision | 92.61% | **93.44%** | 90.87% | 90.76% | 91.26% | 88.96% |
| Recall | 84.04% | 82.99% | 93.88% | 89.42% | **95.06%** | 73.87% |
| F1-score | 86.33% | 86.35% | 90.78% | 88.91% | **92.19%** | 78.97% |
| Accuracy | 18.26% | 16.17% | **26.94%** | 17.46% | 26.39% | 4.81% |

Note: Only documents covered by the specific data source were included in the calculation of precision, recall, F1-score.

With respect to the diachronic changes, the patterns on the WoS gold standard are similar to those on the Scopus gold standards. The coverage, recall, and F1-score of all data sources increase on recent publications, and the gaps between data sources become narrower (Fig.8 and 9). Two versions of COCI outperform other data sources in terms of precision (Fig. 9c). Different from the Scopus gold standards, the accuracy and precision fluctuate on the WoS gold standard. They even witness decreases for publications in recent years (Fig. 9a and 9c). By manually analyzing the raw data, we found that the metadata of unpublished references (in press) is absent from WoS's export results. They were merely labeled as "in press" and would not be updated after publishment. This situation became more common in later years. Moreover, as suggested by a previous study (van Eck & Waltman, 2017), we also found the problem of incorrect references. WoS might provide incorrect metadata for references or even totally irrelevant references. Therefore, the experiment on the WoS gold standard may underestimate the precision of all data sources, which also significantly affects the accuracy. Nonetheless, the relative performances of databases are still informative. While Dimension still has the largest coverage during the whole period, NIH-OCC consistently excels other data sources in terms of recall and F1-score.

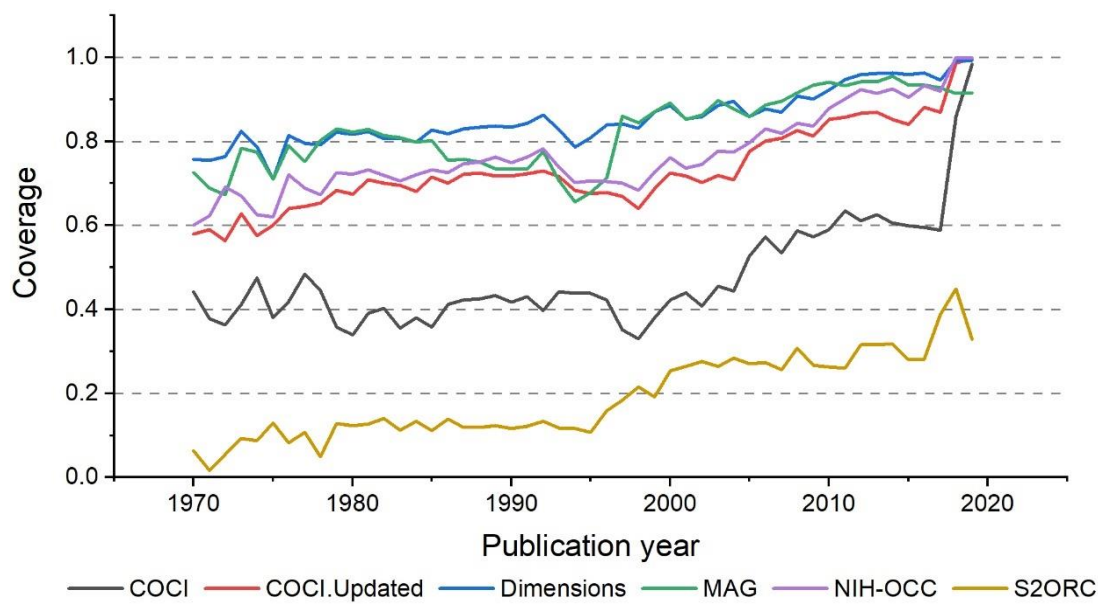

**Fig. 8.** Temporal changes in the coverage of different data sources on the WoS gold standard.

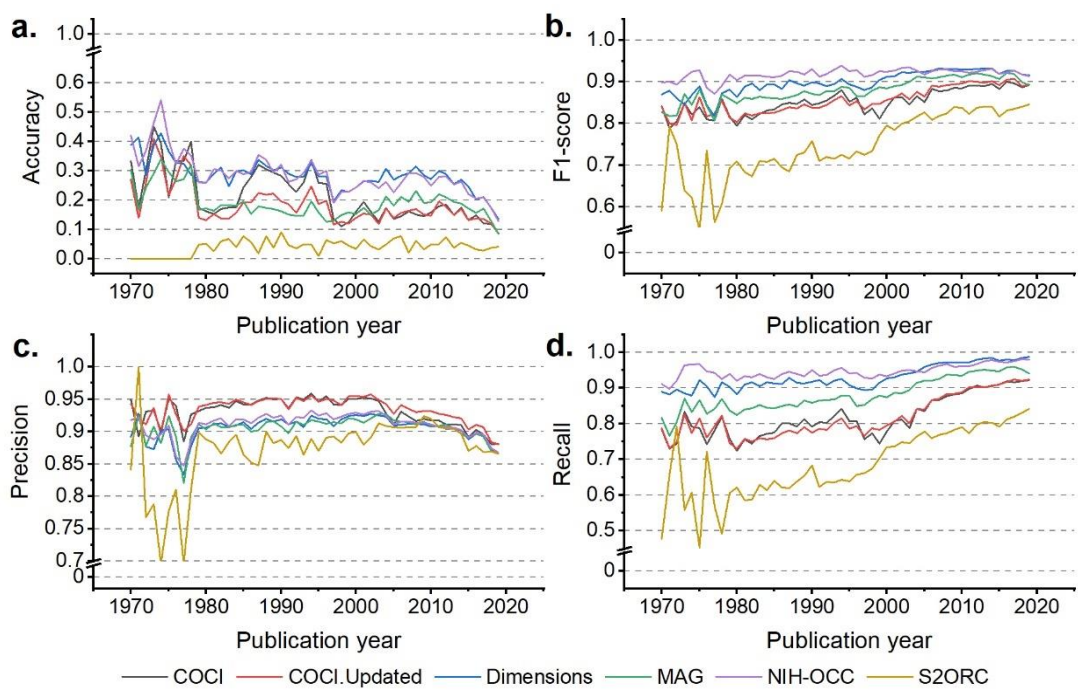

**Fig. 9**. Temporal changes in the accuracy, precision, recall, and F1-score of different data sources on the WoS gold standard.

## Discussion

**Accessibility of the data sources**

The accessibility is yet another major factor to be considered for researchers who decide to perform large-scale citation analyses based on bibliographic data sources. Data sources that support bulk download and API access can significantly reduce the burden of data collection. Table 9 lists some characteristics of the data sources analyzed in this study. Among them, NIH-OCC is the only data source that does not require an application form and supports bulk download and API (bulk) access. Access to COCI is also convenient, one can download the latest release of its entire dataset within an hour. On the other hand, users must submit an application form or even a research proposal to request access to Dimensions, MAG (the latest version), and S2ORC. This procedure may take weeks to complete.

It is worth noting that Microsoft delivers the latest version of MAG through the Microsoft Azure platform and provides computational functions through Azure Databricks, a commercial big data computing platform. Though applying for MAG is free, it still costs money to store, compute, and download the dataset. Users may choose archived versions of MAG, which is available on archive.org at no cost[6]. Unfortunately, during the peer-review process of this paper, Microsoft announced that MAG would be retired by December 31, 2021. At that time, OpenAlex may be a promising replacement yet to be released[7].

---

[6] https://archive.org/search.php?query=creator%3A%22Microsoft+Academic%22
[7] https://openalex.org/

Finally, collecting data from Dimensions is the most time-consuming as it only allows downloading its data in batches up to 500. Nonetheless, such an effort is worthwhile because both the coverage and quality of the Dimensions data are satisfying.

**Table 9**
Accessibility of the data sources.

|  | Require application | Bulk download | API access | API bulk access | License |
| --- | --- | --- | --- | --- | --- |
| COCI | No | Yes | Yes | No | CC0 |
| Dimensions | Yes | No | Yes | Yes, 500 at a time | N/A |
| MAG | Yes/No* | Yes | No | No | ODC-By |
| NIH-OCC | No | Yes | Yes | Yes, 1000 at a time | CC0 |
| S2ORC | Yes | Yes | No | No | CC BY-NC 2.0 |

* Application is required to obtain the latest release of MAG, but archived versions are available on archive.org.

While all bibliographic data sources employed in this study are freely available, applying a dichotomy of "open data" and "non-open data" is still valuable for researchers. According to the definition of "open" given by Open Knowledge Foundation[8], open data should allow "freely use, modify, and share by anyone for any purpose". COCI, NIH-OCC, and the archived versions of MAG can be classified as open data sources as they meet all the above characteristics, indicated by their data licenses. However, both Dimensions and S2ORC require users to submit applications beforehand and forbid commercial purposes. Dimensions also restricts the distribution of its data to any other person. Since not anyone can use the two datasets for any purpose, they cannot be considered as open.

**Comparison with previous studies**

Previous studies on comparing databases primarily focus on the coverage. MAG is considered an excellent alternative for citation analysis as it covers more documents than Scopus and WoS (Harzing 2016; Harzing and Alakangas 2017; Martín-Martín et al. 2020). According to our results, although 89.16% of documents in other data sources also exist in MAG, their references suffer from data incomplete problem. This problem is even more serious in COCI, which is consistent with the research of Visser et al. (2021). Therefore, caution should be taken when adopting MAG or COCI alone as the data source. Visser et al. (2021) also pointed out that missing citation links is a significant problem in Dimensions. However, Dimensions provides high-quality data with large coverage in our study, both its precision and recall are satisfying. One possible reason is that documents in PubMed are mostly from biomedical and life science, while Visser et al. (2021) perform their comparisons based on the entire Scopus and Dimensions database. In addition, Dimensions is a fast-growing database that continuously expands its coverage (Herzog et al. 2020). The data incomplete issue may have been greatly alleviated.

**Implications**

The implications of this research are mostly practical. With an increasing number of freely available bibliographic databases, researchers may have both interest in and concerns about substituting the expensive commercial databases with the free ones. However, to what extent these data sources resemble the commercial databases they used to employ should be investigated beforehand. Our results on the PubMed Baseline show that Dimensions and NIH-OCC are the ideal data sources to retrieve PubMed-to-PubMed citations at no cost, with high coverage and accuracy. Researchers can also combine multiple data sources to generate a more comprehensive dataset, depending on the sensitivity of their studies. For instance, integrating data sources with high precision but low recall is an effective way to enhance the overall recall without undermining the precision.

---

[8] https://opendefinition.org/

Although we mainly focused on the citations within PubMed, the comparison between different data sources was performed on a large scale. We also established gold standards based on authentic data sources and constructed representative samples to investigate the completeness and correctness of citations. The relative performances of databases are stable across different gold standards. Therefore, our results can provide valid references for researchers to choose the appropriate free bibliographic databases for their studies, especially for those based on PubMed documents.

**Limitations**

This study has several limitations. First, without access to the full data of Scopus and WoS (i.e., an in-house version), the referencing relationships in the second and third gold standards were obtained from the websites of Scopus and WoS. Due to the rate limit, we constructed the Scopus and WoS gold standards based on random samples of 50,000 documents. The samples are representative, but the generality can be further improved by using all citation data from Scopus and WoS. Second, gold standards are not perfect. False positives (incorrect references) and false negatives (incomplete references) may occur due to the errors in commercial databases and our conservative matching criteria, respectively. This may lead to underestimation of the data sources' performance. However, the errors in gold standards are rare and the relative performances of databases are stable. The results in the current work are still valid.

**Conclusions**

To conclude, this study aims to help researchers to select appropriate free bibliographic data sources for citation-based studies on PubMed. We compare five freely available data sources in terms of their ability to provide high-quality citation data for PubMed documents. Dimensions turns out to be the most comprehensive data source that provides references for 62.4% of PubMed documents, outperforming the official NIH-OCC dataset. In addition, over 90% of citation links provided by other data sources can be found in Dimensions. The correlation between citation counts of documents in different data sources is strong, suggesting high consistency on citation ranks across data sources.

The results also show that while the coverage is an important factor in choosing data sources, it is still necessary to evaluate the correctness and completeness of the citation data. On the one hand, large coverage does not guarantee high accuracy, e.g., MAG. On the other hand, a data source with smaller coverage may also have competitive performance, e.g., COCI. Overall, NIH-OCC and Dimensions achieve the best results. While NIH-OCC performs better on the PubMed documents preserved by NLM (the first gold standard), Dimensions is still competitive and even outperforms NIH-OCC on a more general population of PubMed documents. Almost all data sources have a precision higher than 90%, but their recall is much lower. All databases have better performances on recent publications than earlier ones. Meanwhile, the differences in performance of data sources have diminished for documents published in recent years.

Taking accessibility into consideration, one can obtain a large volume of high-quality citation data from NIH-OCC with ease and use it for any purpose, while it takes more effort to collect data from Dimensions. Our results show that both databases provide high-quality referencing relationships and have a large coverage, which makes them suitable data sources for augmenting PubMed. It is also beneficial for researchers to combine citations from multiple data sources to build a more comprehensive dataset.

**Acknowledgments** This study was partially funded by the National Natural Science Foundation of China (NSFC) Grant Nos. 71921002 and 71804135. We sincerely thank Dr. Silvio Peroni and anonymous reviewers for their valuable comments. Zhentao Liang would also like to give special thanks to his fiancée Lishan Mai and family for their supports throughout the hard times.